# Macroscopic Virtual Particles Exist

Günter Nimtz

II. Physikalisches Institut, Universität zu Köln

**Abstract:**

*Virtual particles expected to occur in microscopic processes as they are introduced, for instance by Feynman in the Quantum Electro Dynamics, as photons performing in an anonymous way in the interaction between two electrons. This note describes macroscopic virtual particles as they appear in classical evanescent modes and in quantum mechanical tunneling particles. Remarkably, these large virtual particles are present in wave mechanics of elastic, electromagnetic, and Schrödinger fields.*

# 1. Introduction

I am going to talk about the classical evanescent modes and the quantum mechanical tunneling solutions. The first are the imaginary solutions of the Helmholtz equation

$$\Delta^2 \phi(x) + (n\,\omega/c)^2\, \phi(x) = 0, \qquad (1)$$

where $\phi$ is a wave function, n is the refractive index, $\omega$ the angular frequency, and c the velocity of light.

Solved by a plane wave ansatz, the wave number k is given by the refractive index

$$k^2 = (n\,\omega/c)^2 = k_0^2\, n^2. \qquad (2)$$

Imaginary solutions of the stationary Schrödinger equation describe the tunneling process,

$$\Delta^2 \Psi(x) + 2m/\hbar^2\,(E - U(x))\Psi(x) = 0. \qquad (3)$$

The wave number is

$$k^2 = 2m/\hbar^2 \, (E - U(x)), \tag{4}$$

where $\Psi$ is a propagating wave, m is the mass of the particle, E the energy of the stationary state, and U(x) the position-dependent potential.

In the following it is shown that the imaginary wave solutions represent virtual wave packets. Both so called evanescent modes and tunneling solutions are described by a negative energy E and an imaginary momentum p. Thus the Einstein energy relation is violated,

$$E^2 = p^2c^2 + (mc^2)^2. \tag{5}$$

Merzbacher [1] pointed out that in order to observe a particle in the barrier it must be localized within a distance $\Delta x \approx 1/\kappa$, where $\kappa$ is the imaginary wavenumber. Hence its momentum $\Delta p$ must be uncertain by

$$\Delta p > \hbar/\Delta x \approx \sqrt{2m(U - E)}. \tag{6}$$

The particle energy E can thus be located in the non-classical region only if it is given an energy (U–E) sufficient to raise it into the classically allowed region. This effect is well known by electronic engineers when they characterize waves in an undersized wave guide.

In the following, evanescent and tunneling modes are explained as virtual particles.

Introducing virtual photons, Feynman wrote [2]: Such an exchanged photon that never appears in the initial or final conditions of the experiment is sometimes called a "virtual photon." Brillouin mentioned in his textbook on *Wave Propagation in Periodic Structures* [3]: The zone structure is completely independent of the special physical meaning of the waves considered, and it must be the same for elastic, electromagnetic, and Schrödinger electronic waves. This statement is presumably valid for all phenomena of fields, which have wave mode solutions. As recently as the seventies of the last century, it was shown that the electromagnetic evanescent modes are virtual particles described by the QED [4-6]. Cargnilia and Mandel have shown that the commutator of field operators between two space-like separated points does not vanish, thereby violating the microscopic causality condition [4, 6].

## 2. Examples of tunneling barriers

*2.1 Total reflection*

The most studied case of evanescent modes is the frustrated total reflection at a double prism [7 - 9]. As shown in Fig 1, some part of the incoming beam under the angle of total reflection is transmitted to the second prism, thus frustrating the total reflection. This effect decreases exponentially with the exponential κ in enlarging the gap d, where θ is an angle larger than that of total reflection. $n_1$ and $n_2$ are the corresponding refractive indices,

$$\kappa = \frac{\omega}{c} \sqrt{(n_1^2/n_2^2)\sin2\theta - 1}. \qquad (7)$$

As outlined in Fig. 1, Sommerfeld explained the double prism as the classical analog of the quantum mechanical tunneling process [10]. Incidentally, in classical physics evanescent modes and thus tunneling are not possible. However, Sommerfeld's statement is not surprising having in mind the mathematical equivalence of the Helmholtz equation (1) and the stationary Schrödinger equation (3). Brillouin has pointed to the general validity of such phenomena in all fields, where wave mechanics holds [3].

Fig. 2 displays the virtual process of photons crossing a double prism [9]. Here, both prisms were obtained by a diagonal cut of a single 40 cm cube of Perspex. The photons couple with the dielectric dipoles of the prisms to form real polaritons [11, 12].

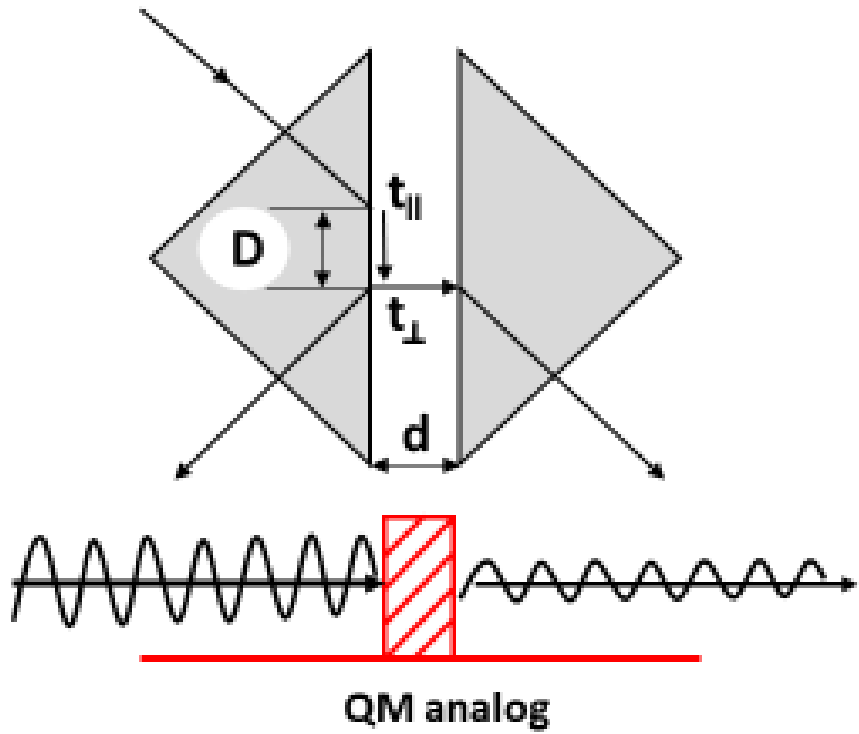

*Fig. 1 Double prism, the optical QM tunneling analog according Sommerfeld [10]. E and U are the particle energy and the potential barrier height. D is the Goos-Hänchen shift along the surface and d the width of the gap between the two prisms [9]. The gap traversal time $t_\perp$ is observed to be zero and $t_\parallel$ represents the effective tunneling time τ. In such a symmetric set-up both the reflected and the transmitted beams are detected at the same time [9].*

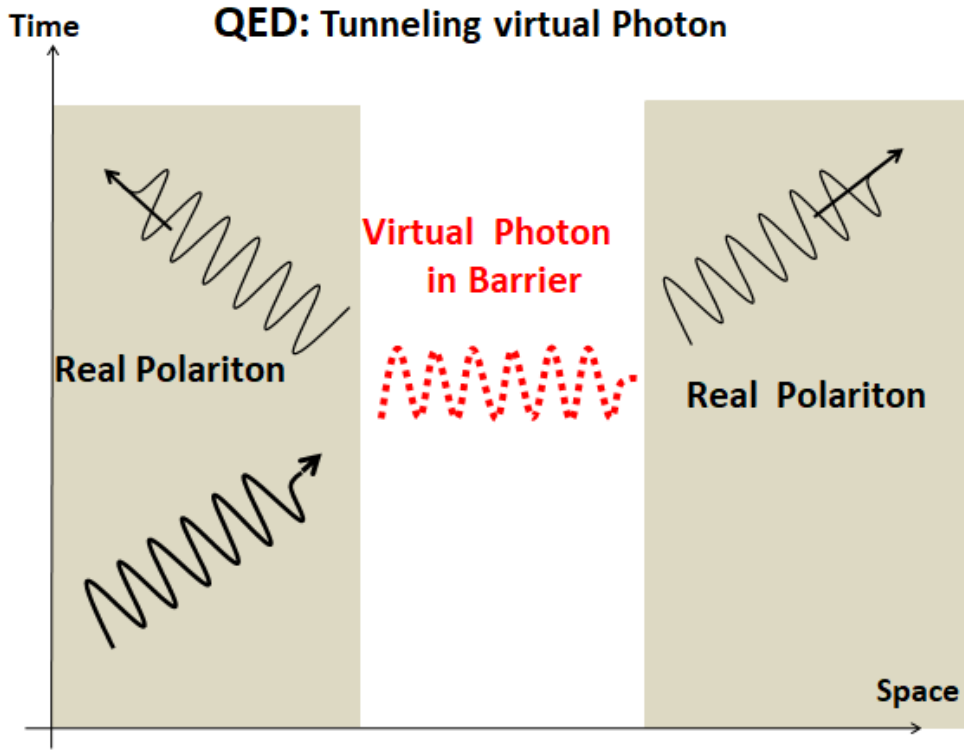

*Fig. 2. Minkowski diagram of the double prism shown in Fig. 1. The length of the virtual photon is about 3 cm [11, 12].*

## 2.2 Lattice structures

Periodically varying potential structures have at distinct wavelengths an increasing reflection with the number of layers. This is due to the partial destructive interference of incoming and reflected waves from the two different dielectric layers. For instance, a plain one-dimensional lattice is built by elementary cells that are given by the two quarter wavelength layers of thickness $d_1$ and $d_2$, one with and the other without a phase shift of $\pi/2$ in reflection so that

$$n_1 d_1 = n_2 d_2 = \lambda_0/4, \tag{8}$$

where $\omega_0 = 2\pi c/\lambda_0$. $\omega_0$ is the forbidden mid-gap angular frequency of the special arrangement. Such a transmission mid-gap is shown in Fig. 4.

The general properties of lattice structures are studied in Refs. [13, 14].

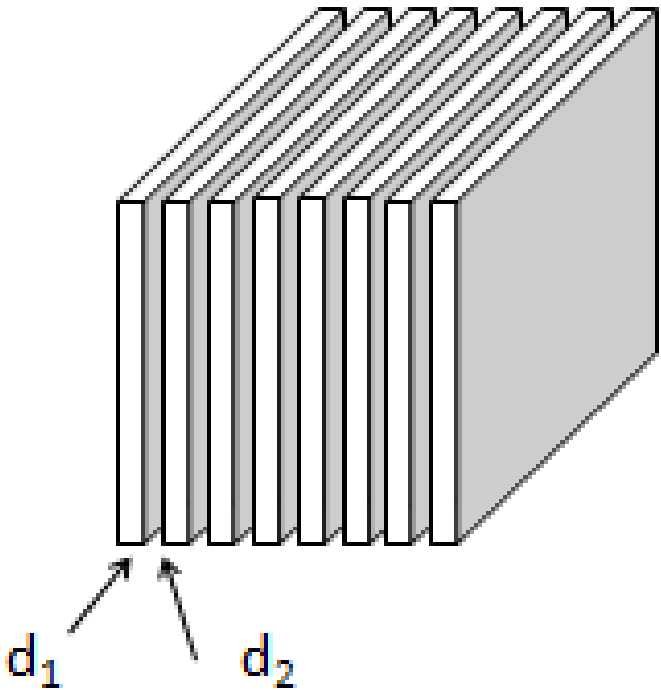

*Fig. 3 . Sketch of a one-dimensional lattice structure of periodical varying layers of two different refractive indices, i.e. field potentials.*

An infrared digital signal tunneling through a lattice is shown in Fig. 4.

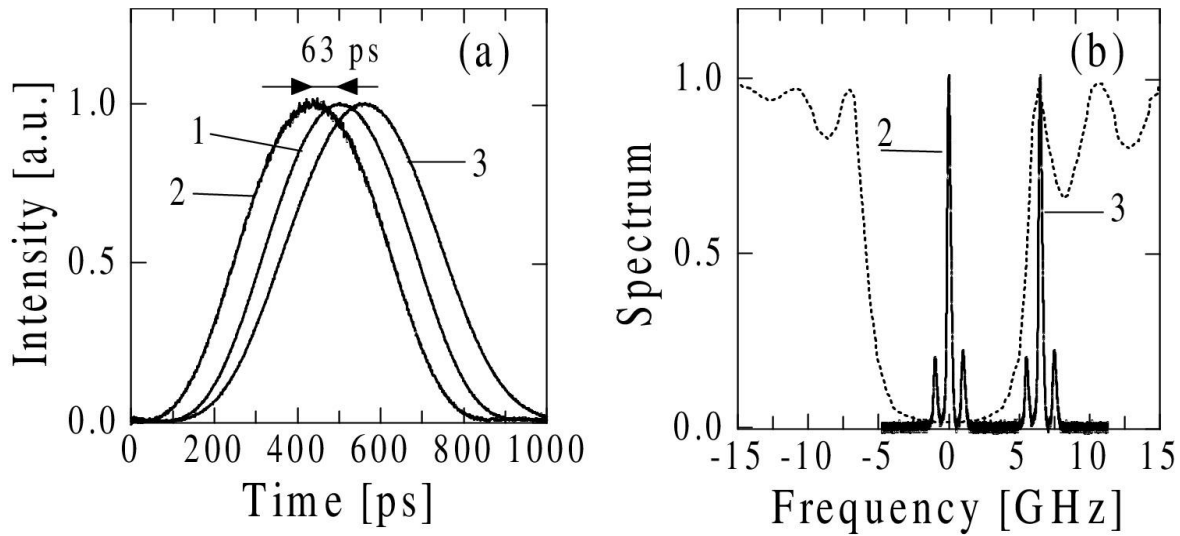

*Fig.4. (a) Transmission of a digit on a carrier of 1.5 µm wave length and (b) the gap structure on a glass fiber lattice [15]. 2 represents the tunneled digit, 1 the not tunneled one on the fiber, and 3 traveled at a frequency above the tunneling regime. All three digits traveled the same distance.*

Electromagnetic wave tunneling experiments were carried out in the wavelength range between µm and cm by more than 5 orders of magnitude, i.e. from infrared up to microwave frequencies. Incidentally, in laser technology, mirrors of dielectric structures are favored over metallic mirrors due to their smaller losses.

## 2.3 Undersized waveguides

Consider microwaves in the frequency range between 1 and 100 GHz, i.e. 30 cm to 300 mm wavelengths, when channeled in metallic guides as sketched in Fig. 5.

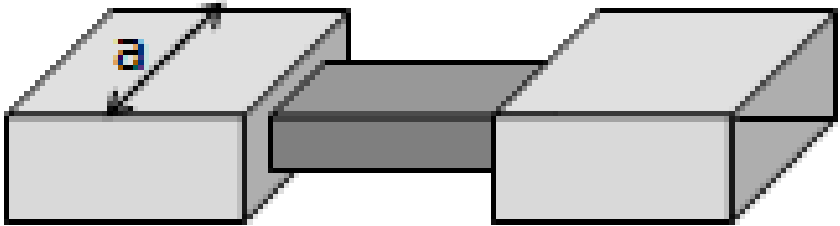

*Fig. 5. A rectangular brass waveguide. Its advantage is the electric polarization of the wave (TE₁₀ mode).*

The ideal metallic waveguide has the relation of the wave vector

$$k = k_0 \sqrt{1 - (\lambda_0/\lambda_{cutoff})^2} = \sqrt{1 - (\omega_0/\omega_{cutoff})^2} \quad (9)$$

$$k = k_0\, n(\omega)$$

where $\lambda_{cutoff} = 2a$, a is the waveguide width of the basic microwave mode $TE_{10}$ [16]. Cutoff means that for $\lambda \geq 2a$ the wave vector is imaginary, the wave has become evanescent. This waveguide relation is analogous to Eq. (4).

With an undersized waveguide acting as a tunnel barrier between normal sized transmission lines, an overall superluminal transmission due to zero time inside the cutoff range was measured by Enders and Nimtz for the first time in 1992 [17, 18]. The measured tunneling time is found as the time spent at the barrier front of opaque barriers, whereas the time inside the barrier is zero [12, 18, 24].

Faster than light (superluminal) tunneling has been confirmed in subsequent photonic experiments, for example in Refs. [19, 20]. One of them was carried out in a single photon experiment with the Hong-Ou-Mandel interferometer, a quantum mechanical technique [21], whereas another example was performed with a digital signal on a glass fiber [20]. At that time neither these studies nor several other theoretical investigations presented a correct interpretation of measured faster-than-light signal tunneling as was shown in Ref. [12]. The violent discussion was based on the question of whether causality is violated. (It is not [12].)

## 2.4 Ultrasound Tunneling

Another surprising observation was acoustic supersonic tunneling. A simple barrier for sound tunneling is performed with a lattice structure like that of Fig. 2. The dielectric layers are substituted by two different metals for example lead (v = 2640 m/s) and aluminum (v = 5080 m/s). Supersonic velocities have been studied at MHz frequencies (22).

A different one-dimensional band gap structure was extensively studied by Robertson et al. [23]. Here, an acoustic periodic hetero structure was constructed from standard ¾ in. PVC pluming pipe. The acoustic impedance was periodically changed by short side branches of the same PVC pluming pipe using standard T-junctions. Such an array has similar acoustic dispersion relations to the simple lattice structure. They tunneled at frequencies near 1 kHz observing velocities roughly three times faster than in the same pipe without the tunneling. Ultrasound tunneling through a 3-dimensional phononic crystal was studied by Yang et al. [24]. The crystal was an fcc array of tungsten carbide beads in water. Sound velocities up 16 times the velocity of sound in water have been measured. Remarkably, the measured elastic field tunneling time corresponded to the observed universal photonic tunneling time [17].

## 3. Summing Up

Virtual interaction was thought to take place in the microcosm. Here we have presented virtual waves with lengths up to meters. The demonstrated examples appeared in the tunneling process. Tunneling was said to be a quantum mechanical phenomenon. However, it is a wave mechanics interference effect and holds for all fields with wave solutions. Another special property is that the tunneling happens in zero time. An interaction time arises only at the barrier entrance [18, 25].

## 4. Acknowledgement

It is a pleasure to acknowledge stimulating conversations with Paul Bruney, who raised pertinent questions on aspects of this work.